\documentclass[twocolumn]{aastex62}

\usepackage{amsmath}
\usepackage{graphicx}
\usepackage{lipsum}
\usepackage{hyperref}
\usepackage{mathtools, nccmath}
\usepackage{longtable}
\usepackage{lineno}

\newcommand\AS[1]{{#1}} 
 
\newcommand\Rcoop[1]{{#1}} 
\newcommand{\msun}{M$_{\odot}$}
\newcommand{\Lsun}{L$_{\odot}$}

\newcommand\rev[1]{{#1}}

\graphicspath{{./}{figures/}}

\submitjournal{AAS Journals}

\shorttitle{JWST/NIRISS AMI observations of WR~137}
\shortauthors{Lau et al.}

\begin{document}

\title{A First Look with JWST Aperture Masking Interferometry (AMI): Resolving Circumstellar Dust around the Wolf-Rayet Binary WR~137 beyond the Rayleigh Limit}

\correspondingauthor{Ryan M.\ Lau}
\email{ryan.lau@noirlab.edu}

\author[0000-0003-0778-0321]{Ryan M.\ Lau}
\affil{NSF’s NOIRLab, 950 N. Cherry Ave., Tucson, AZ 85719, USA}

\author[0000-0001-9315-8437]{Matthew  J.\ Hankins}
\affil{Arkansas Tech University, 215 West O Street, Russellville, AR 72801, USA}

\author[0000-0002-9723-0421]{Joel Sanchez-Bermudez}
\affil{Instituto de Astronomía, Universidad Nacional Autónoma de México, Apdo. Postal 70264, Ciudad de México 04510, Mexico}
\affil{Max-Planck-Institut für Astronomie, Königstuhl 17, D-69117 Heidelberg, Germany}

\author[0000-0002-1536-7193]{Deepashri Thatte}
\affiliation{Space Telescope Science Institute, 3700 San Martin Drive, Baltimore, MD 21218, USA}

\author{Anthony Soulain}
\affil{Sydney Institute of Astronomy (SIfA), School of Physics, The University of Sydney, NSW 2006, Australia}

\author[0000-0001-7864-308X]{Rachel A. Cooper}
\affiliation{Space Telescope Science Institute, 3700 San Martin Drive, Baltimore, MD 21218, USA}

\author[0000-0003-1251-4124]{Anand Sivaramakrishnan}
\affiliation{Space Telescope Science Institute, 3700 San Martin Drive, Baltimore, MD 21218, USA}
\affiliation{Astrophysics Department, American Museum of Natural History, 79th Street at Central Park West, New York, NY 10024}
\affiliation{Department of Physics and Astronomy, Johns Hopkins University, 3701 San Martin Drive, Baltimore, MD 21218, USA}

\author{Michael F. Corcoran}
\affiliation{CRESST II and X-ray Astrophysics Laboratory NASA/GSFC, Greenbelt, MD 20771, USA}
\affiliation{Institute for Astrophysics and Computational Sciences, The Catholic University of America, 620 Michigan Ave., N.E. Washington, DC 20064, USA}

\author{Alexandra Z. Greenbaum}
\affiliation{IPAC, California Institute of Technology, 1200 E. California Boulevard, Pasadena, CA, 91125, USA}

\author{{Theodore R. Gull}}
\affiliation{Space Telescope Science Institute, 3700 San Martin Dr., Baltimore, MD 21218, USA}
\affiliation{Code 667, NASA/GSFC, Greenbelt, MD 20771}

\author{Yinuo Han}
\affil{Institute of Astronomy, University of Cambridge, Madingley Road, Cambridge CB3 0HA, UK}

\author[0000-0003-4870-5547]{Olivia C.\ Jones}
\affiliation{UK Astronomy Technology Centre, Royal Observatory, Blackford Hill, Edinburgh, EH9 3HJ, UK}

\author[0000-0001-7697-2955]{Thomas Madura}
\affiliation{Department of Physics and Astronomy, San Jose State University, San Jose, CA, USA}


\author{Anthony F.~J.~Moffat}
\affil{Département de Physique, Université de Montréal, C.P. 6128, succ. centre-ville, Montréal (Qc) H3C 3J7, Canada; and Centre de Recherche en Astrophysique du Québec, Canada}

\author{Mark R.~Morris}
\affil{Department of Physics and Astronomy, University of California, Los Angeles, 430 Portola Plaza, Los Angeles, CA 90095-1547, USA}

\author{Takashi Onaka}
\affil{Department of Astronomy, School of Science, University of Tokyo, 7-3-1 Hongo, Bunkyo-ku, Tokyo 113-0033, Japan}

\author{Christopher M. P. Russell}
\affil{Department of Physics and Astronomy, Bartol Research Institute, University of Delaware, Newark, DE 19716 USA}

\author[0000-0002-2806-9339]{Noel D. Richardson}
\affiliation{Department of Physics and Astronomy, Embry-Riddle Aeronautical University, 3700 Willow Creek Rd, Prescott, AZ 86301, USA}

\author{Nathan Smith}
\affiliation{Steward Observatory, 933 North Cherry Avenue, Tucson, AZ 85721, USA}

\author[0000-0001-7026-6291]{Peter Tuthill}
\affiliation{School of Physics, The University of Sydney, NSW 2006, Australia}

\author{Kevin Volk}
\affiliation{Space Telescope Science Institute, 3700 San Martin Drive, Baltimore, MD 21218, USA}

\author[0000-0001-9754-2233]{Gerd Weigelt}
\affil{Max-Planck-Institut für Radioastronomie, Auf dem Hügel 69, 53121 Bonn, Germany}

\author[0000-0002-8092-980X]{Peredur M.~Williams}
\affil{Institute for Astronomy, University of Edinburgh, Royal Observatory, Edinburgh EH9 3HJ, UK}




\begin{abstract}

We present infrared aperture masking interferometry (AMI) observations of newly formed dust from the colliding winds of the massive binary system Wolf-Rayet (WR) 137 with JWST using the Near Infrared Imager and Slitless Spectrograph (NIRISS). NIRISS AMI observations of WR~137 and a point-spread-function calibrator star, HD~228337, were taken using the F380M and F480M filters in 2022 July and August as part of the Director's Discretionary Early Release Science (DD-ERS) program \#1349. Interferometric observables (squared visibilities and closure phases) from the WR~137 ``interferogram'' were extracted and calibrated using three independent software tools: ImPlaneIA, AMICAL, and SAMpip. The analysis of the calibrated observables yielded consistent values except for slightly discrepant closure phases measured by ImPlaneIA. Based on all three sets of calibrated observables, images were reconstructed using three independent software tools: BSMEM, IRBis, and SQUEEZE. All reconstructed image combinations generated consistent images in both F380M and F480M filters. The reconstructed images of WR~137 reveal a bright central core with a $\sim300$ mas linear filament extending to the northwest. 
A geometric colliding-wind model with dust production constrained to the orbital plane of the binary system and enhanced as the system approaches periapsis provided a general agreement with the interferometric observables and reconstructed images. Based on a colliding-wind dust condensation analysis, we \rev{suggest} that dust formation within the orbital plane of WR~137 is induced by enhanced equatorial mass-loss from the rapidly rotating O9 companion star, whose axis of rotation is aligned with that of the orbit.

\end{abstract}


\section{Introduction}
\label{sec:intro}

Classical Wolf-Rayet (WR) stars, the descendants of massive OB-type stars, are characterized by high luminosities ($L_*\gtrsim10^5$ \Lsun), hot surface temperatures ($T_{\star}\gtrsim$ 40000 K), and fast, powerful winds ($v_{\infty}\gtrsim$ 1000 km s$^{-1}$, $\dot{M}$ $\gtrsim 10^{-5.5}$ \msun~yr$^{-1}$; \citealt{Crowther2007,Hamann2019,Sander2019}). 
The high \rev{luminosities} and intense UV radiation produced by WR stars may present an inhospitable environment for dust grains; however, infrared (IR) observations have demonstrated that a subset of WR \rev{stars exhibits} dust production \cite[e.g.,][]{Allen1972, Williams2019,Lau2020}. These dust-producing objects \rev{have enhanced carbon abundances} (WC stars) and come in two categories: ones that continuously produce dust and those that produce dust in an episodic fashion \citep{Williams1987}. For the majority of these objects, the dust-formation process is thought to be tied \rev{to their binary nature}, where a collision between the WC star wind and that of an OB-type companion produces high densities and efficient cooling that favor dust formation \cite[e.g.,][]{Usov1991}. However, the detailed physics of colliding-wind dust production is not well understood, and much of our present knowledge has been gleaned from periodic dust producers whose dust formation episodes \rev{have} been directly tied to the \rev{binary orbit}. The \rev{archetypical} example of this is WR~140, which produces dust like clockwork during its periastron passage every 7.93 years \citep{Williams2009,Lau2022}. 

WR 137 (also known as HD~192641) is another well-known dust producing WC binary which shows repeating dust formation events \citep{Williams1985,Williams2001}. It is a confirmed binary consisting of a WC7 star and O9e type companion \citep{St-Louis2020}, and has a measured orbital period of \rev{13.1} years \citep{Lefevre2005}.
The IR light-curve of WR~137 exhibits high-amplitude ($\gtrsim1$ mag) IR brightening episodes with a similar cadence to the orbital period of the binary \citep{Williams2001}, which suggests that dust formation is orbitally modulated \rev{as in} WR~140. 
However, previous images of WR~137 taken in the near-IR by the \textit{Hubble Space Telescope} (HST) present compact, ``jetlike'' dust emission \citep{Marchenko1999} that does not resemble any of the other known dust-forming WC binary systems \citep{Tuthill1999,Monnier2007,Lau2020WR112}. Investigating the connection between WR~137's dust formation and its orbital properties has been challenging due to the compact morphology of its extended dust emission.
Observations with high spatial resolution, high imaging contrast, and high sensitivity at mid-IR wavelengths are therefore essential for resolving and characterizing the nature of the faint extended dust emission around WR~137.

\begin{figure*}[t]
    \includegraphics[width=0.95\linewidth]{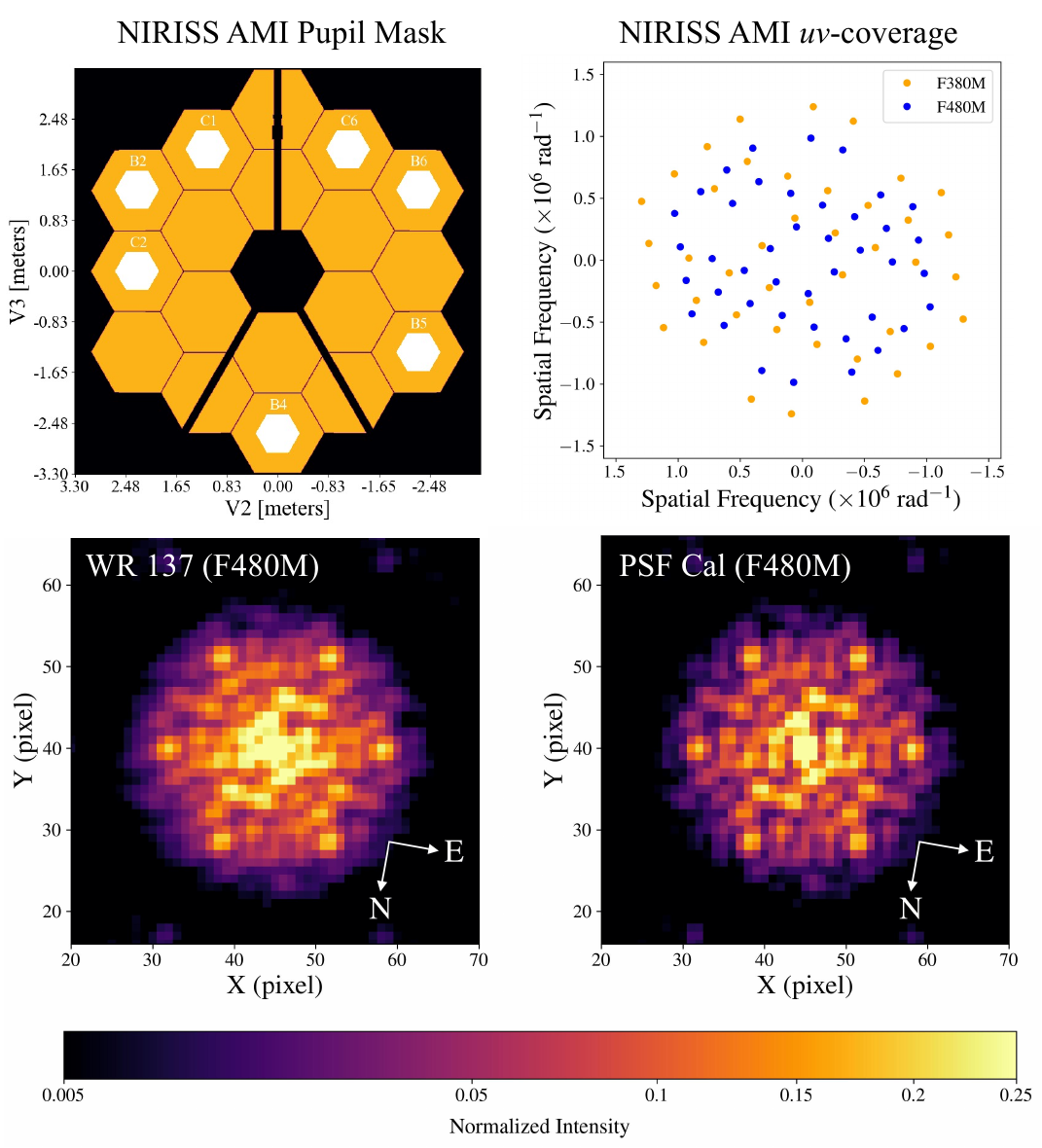}
    \caption{(\textit{Upper Left}) The 7-hole non-redundant NIRISS AMI pupil mask (white hexagons) overlaid on an outline of the JWST primary mirror. (\textit{Upper Right}) The uv-plane coverage of the non-dithered F380M and F480M observations of WR~137. (\textit{Bottom Left}) \rev{$50\times50$ pixel cutout centered on the}  interferogram pattern produced by the 7-hole non-redundant mask \rev{on the $80\times80$ pixel SUB80 subarray} for WR~137 and (\textit{Bottom Right}) the PSF calibrator star\rev{, both with} the F480M filter. The images displayed are a median of frames that have gone through both stage 1 and 2 processing, and are shown with a normalized power-law stretch (index = 0.5). Each pixel corresponds to 65 mas and the colorbar corresponds to intensity normalized to the peak pixel value. Although subtle, the WR 137 interferogram pattern appears to have a lower contrast than the PSF calibrator interferogram. 
    This is the result of the structure other than the bright central compact source and therefore indicates the presence of resolved emission around WR137.}
    \label{fig:interferogram}
\end{figure*}

As part of Director's Discretionary Early Release Science (DD-ERS) program \#1349, WR~137 was selected as a target to investigate colliding-wind dust formation and to demonstrate the \rev{scientific} potential of the Aperture Masking Interferometry (AMI) mode of NIRISS. 
The 7-hole, non-redundant AMI pupil mask is shown in Fig.~\ref{fig:interferogram} (\textit{Upper Left}) overlaid on an outline of the JWST primary mirror.
Although the usage of the non-redundant mask (NRM) that enables AMI\footnote{Also referred to as Sparse Aperture Masking (SAM).} on NIRISS blocks $\sim85\%$ of the incoming light \citep{Artigau2014}, AMI techniques provide angular resolution that is $\sim$2$\times$ higher compared to traditional imaging (i.e.~$\sim0.5\,\lambda/D$). This facilitates studies of small-angular-scale dust features near a bright point source.

Previously, AMI has been used on large, ground-based telescopes to study the size and morphology of several dusty WR stars at IR wavelengths \cite[see e.g.,][]{Tuthill1999, Monnier2007, Rajagopal2007, Hankins2016}. 
JWST/NIRISS AMI operating at a temperature of 40 K brings advantages of space-based interferometry at these wavelengths: about an order of magnitude better fringe phase measurements, which are sensitive to the point-antisymmetrical component of the target, two or three orders of magnitude better fringe amplitude calibration that track symmetrical target structure, and much lower thermal background than ground-based high-resolution telescopes \citep{Sivaramakrishnan2009,Sivaramakrishnan2023}. Notably, revealing diffuse extended components such as disks or rings is challenging from the ground because fringe amplitudes are often corrupted by atmospheric scintillation.
JWST has therefore opened a new window on AMI observations from space and \rev{enables} investigations that require high spatial resolution at mid-IR wavelengths and greater sensitivity to extended structures.

In this paper, we present JWST/NIRISS AMI observations of WR~137 using the F380M and F480M filters. 
The timing of the observations was planned to align with active dust formation from WR~137 based on its IR light curve and known spectroscopic orbit (\citealt{Williams2001,Lefevre2005,Peatt2023}). 
In Sec.~\ref{sec:Obs}, we describe the observations, preparation, data reduction, and the procedures used to extract interferometric observables from the NIRISS AMI data.  
In Sec.~\ref{sec:RA}, we present reconstructed images of WR~137 using three different image reconstruction software packages which we then analyze and compare with geometric dust models used to study other dust-forming WC binaries (e.g.,~\citealt{Han2022}). Lastly, in Sec.~\ref{sec:discussion}, we discuss WR~137's dust morphology and the possible influence of its rapidly rotating companion star on conditions for dust formation via wind-wind collision.

\begin{deluxetable*}{p{2.4cm}p{1.6cm}p{2.3cm}p{2.6cm}p{2.3cm}p{2.3cm}p{2.3cm}}
\tablecaption{Summary of Observations}
\tablewidth{0pt}
\tablehead{Obs. Date & Object & Filter & Obs. Type & NGROUPS & NINTS & Dither Pattern}\
\startdata
2022-07-15 13:37:18.246 & WR~137 & F480M & NRM/SUB80 & 4 & 1600 & Stare \\
2022-07-15 13:37:17.634 & WR~137 & F380M & NRM/SUB80 & 2 & 2720 & Stare \\  
2022-07-15 12:12:59.178 & HD~228337 & F480M & NRM/SUB80 & 7 & 1020 & Stare \\
2022-07-15 12:20:57.815 & HD~228337 & F380M & NRM/SUB80 & 2 & 2720 & Stare \\
\hline
2022-07-13 12:43:39.770 & WR~137 & F480M & NRM/SUB80  & 4 & 400 & 4-point \\
2022-07-13 12:43:37.531 & WR~137 & F380M & NRM/SUB80  & 2 & 680 & 4-point \\
2022-08-09 17:33:05.182 & WR~137 & F480M & NRM/SUB80  & 4 & 400 & 4-point \\
2022-08-09 17:33:00.515 & WR~137 & F380M & NRM/SUB80  & 2 & 680 & 4-point \\
2022-08-09 17:51:33.502 & HD~228337 & F480M & NRM/SUB80 & 7 & 255 & 4-point \\
2022-08-09 17:33:45.279 & HD~228337 & F380M & NRM/SUB80 & 2 & 680 & 4-point \\
\enddata
\tablecomments{Summary of the JWST/NIRISS AMI observations of WR~137 and the PSF calibrator HD~228337. All AMI observations were taken using the NIRISS non-redundant mask (NRM) in the SUB80 subarray, which is standard for the AMI mode. NGROUPS and NINTS correspond to the number of groups per integration and the number of integrations per exposure, respectively. Observations using two different dither patterns, stare and 4-point, were performed with identical total exposure times. Note that a duplicate set of dithered data of WR~137 exists because PSF calibrator observations that were planned to follow the 13 July observations of WR~137 were skipped due to a mirror ``tilt event.'' The set of WR~137 and PSF calibrator observations were successfully completed on 9 August. } 
\label{tab:Obs}
\end{deluxetable*}

\section{Observations and Data Reduction}
\label{sec:Obs}

\subsection{JWST/NIRISS AMI Observations, Preparation, and Data Reduction}
\label{sec:reduction}

JWST observations of WR 137 were carried out as part of DD-ERS program \#1349 (PI: R.~Lau) and were obtained using the AMI mode of the NIRISS instrument \citep{Doyon2023,Sivaramakrishnan2023}. Observations of WR~137 and a PSF calibrator star (HD~228337) were taken using the F380M ($\lambda_\mathrm{pivot}=3.825$ $\mu$m; $\Delta \lambda=0.205$ $\mu$m) and F480M ($\lambda_\mathrm{pivot}=4.815$ $\mu$m; $\Delta \lambda=0.298$ $\mu$m) filters on 2022 July 13 and 15, and 2022 August 9.

The NIRISS AMI observations used the standard AMI parameters with the ``NISRAPID'' readout pattern and the SUB80 ($80\times80$ pixel) subarray, where the NIRISS detector plate scale is 65 mas/pixel and the readout time is 75.44 ms. The total\rev{/effective} exposure times of the F480M and F380M observations of WR~137 were 10.6\rev{/8.0} min and 11.2\rev{/6.8} min, respectively. The total exposure time of the PSF calibrator is identical to that of WR~137. Details of the observations can be found in Table~\ref{tab:Obs}. 
For both WR~137 and the PSF calibrator, target acquisition was used on the same target as the observation in the ``AMIBRIGHT'' acquisition mode with the F480M filter and 5 groups in the NISRAPID readout pattern.
\rev{NIRISS AMI observations also allow for direct imaging using the ``CLEARP'' aperture with the same filters and subarray used for the AMI exposures.  Direct F380M and F480M images of WR~137 were therefore obtained with the AMI exposures on 2022 July 15. The total/effective exposure times of the direct F480M and F380M observations were 39.7/31.7 sec and 39.5/24.1 sec, respectively. However, as expected due its brightness, the PSF core of WR~137 was saturated in both filters and was therefore not used for the analysis in this paper.}

In preparation for the observations, simulated NIRISS AMI data products of WR~137 and PSF calibrator HD~228337 were generated to ensure the success of the AMI observing strategy. The Multi-Instrument Ramp Generator (MIRaGe; \citealt{Hilbert2022}) was used to simulate raw NIRISS AMI data in the F380M and F480M filters. MIRaGe simulations utilized the Astronomer's Proposal Tool (APT) file for DD-ERS \#1349, a simulated sky scene of WR~137, and a source catalog with the flux densities of HD~228337 to generate raw exposures identical to the observing configuration specified in the APT file. The simulated raw exposures were then processed through the same procedures \rev{described further in this section as were performed} on the real data. The simulated reduced data products demonstrate that the NIRISS AMI observing configuration described above is capable of detecting the predicted extended dust emission near WR~137 \citep{Marchenko1999,Williams2001}.

Due to WR~137's brightness in the mid-IR ($F_{3.8\mu\mathrm{m}} \gtrsim 1$ Jy; \citealt{Williams2001}), it was important that the number of groups used per integration (NGROUPS) in the observations be specified to avoid non-linearity effects. Charge migration, also known as the ``brighter-fatter effect'' \citep{Hirata2020}, is of particular concern for AMI data and occurs due to the buildup of a transverse electric field as charge accumulates in sufficiently illuminated pixels (e.g. \citealt{Coulton2018}). This effect can significantly impact AMI data given that the core of the PSF contains critical information about the structure of the source.

\AS{Commissioning tests indicated that charge migration between the brightest pixel at the center of an image of an unresolved star and its eight neighboring pixels  remained below 1\% as long as the peak pixel's accumulated up-the-ramp charge remained below 30,000 electrons, the nominal saturation limit for AMI mode \citep{Sivaramakrishnan2023}.} 
\Rcoop{Sampling up the ramp with sufficient NGROUPS allows count rate in the peak pixel and surrounding pixels to be examined as a function of exposure time by processing sections of each integration in an exposure as if they were independent exposures.
Observations of WR~137 with NGROUPS = 4 and the PSF calibrator star with NGROUPS = 7 were examined in this way and found to display non-linearity due to charge migration of at most 2\% at the highest signal level reached. The impact of charge migration on our dataset is therefore negligible. }

Two sets of observations with identical exposure times were obtained using different dither patterns in order to investigate the optimal dither pattern for the NIRISS AMI mode. The two modes used were a non-dithered `stare' mode and a 4-point dither pattern with 4 primary dithers. Although further work is planned for a more detailed investigation of the dither vs. stare mode, the stare observations are used for the analysis in this paper because of larger uncertainties in the interferometric observables extracted from the dithered observations.  
Figure~\ref{fig:interferogram} (\textit{Upper Right}) shows the uv-plane coverage of the non-dithered F380M and F480m observations and presents the spatial frequencies sampled by the non-redundant mask.
A duplicate set of dithered data of WR~137 exists (Tab.~\ref{tab:Obs}) because PSF calibrator observations that were supposed to be taken following the 13 July observations of WR~137 were skipped due to a mirror ``tilt event'' \citep{Rigby2023}. The set of WR~137 and PSF calibrator observations using the 4-pt primary dithers \rev{was} rescheduled for 9 August and successfully completed.

\begin{figure*}[t!]
    \includegraphics[width=0.98\linewidth]{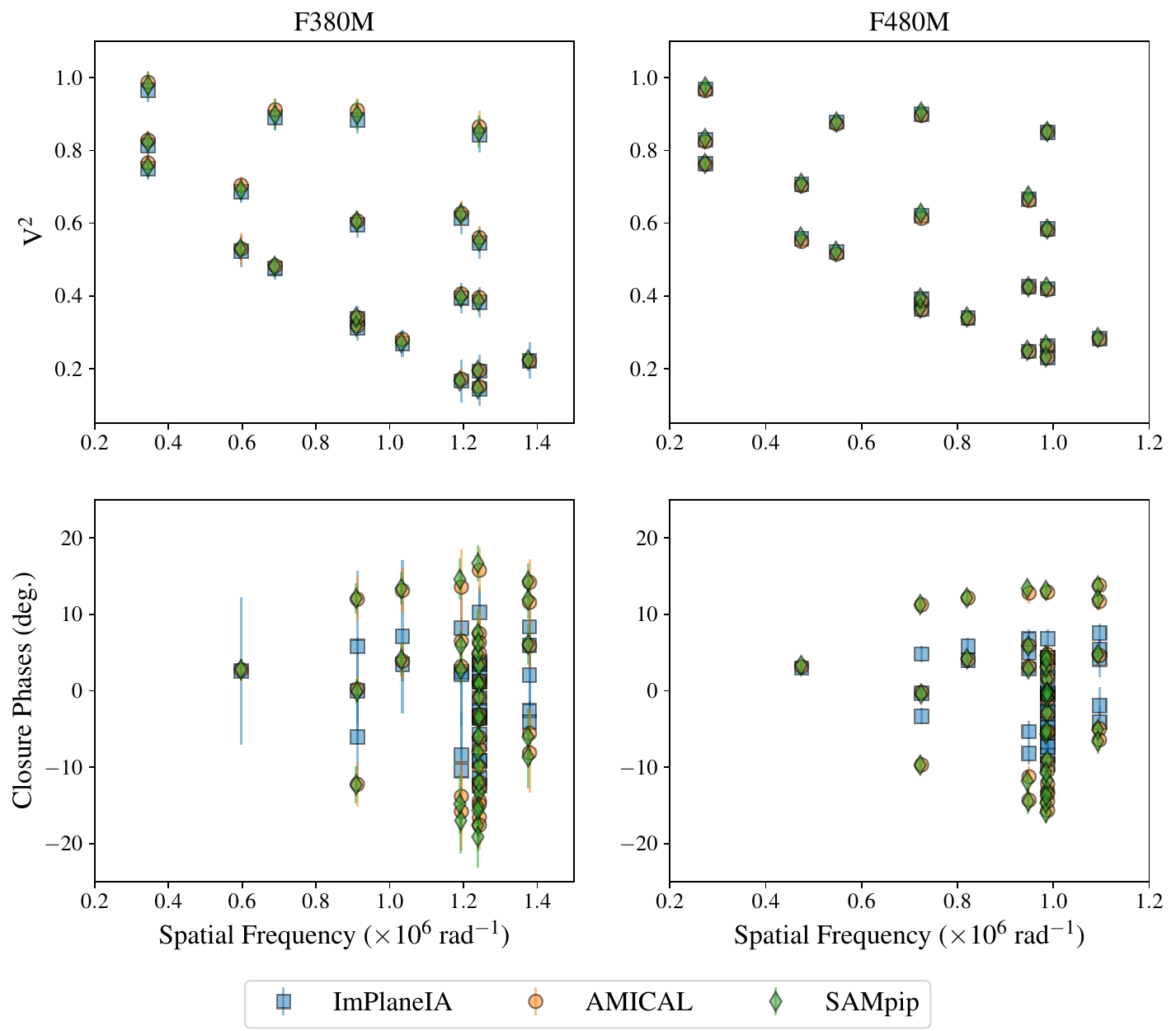}
    \caption{Calibrated Interferometric observables extracted from the WR137 observations using ImPlaneIA, AMICAL, and SAMpip. These data include calibrated squared visibilities (V$^2$, \textit{top}) and closure phases (\textit{bottom}) for the F380M (\textit{left}) and F480M (\textit{right}) filters. }
    \label{fig:IOplots}
\end{figure*}

The observations of WR~137 and the PSF calibrator were reduced using version 1.11.2 of the standard JWST science calibration pipeline with version 11.17.2 of the Calibration Reference Data System (CRDS) for stage 1 and 2 processing. The raw data were first passed to the \textit{calwebb\_detector1} routine (stage 1),
a series of detector level processing steps which take individual integrations and produce a corrected count rate image. The data were then passed to the \textit{calwebb\_image2} routine (stage 2),
which performs instrumental corrections and calibrations on the count rate images to produce fully calibrated exposures.

Figure~\ref{fig:interferogram} (\textit{Bottom}) presents \rev{a cutout of} the ``interferogram'' patterns from the calibrated, non-dithered F480M exposures of WR~137 and the PSF calibrator.
\rev{Since the mask holes are not obstructed by and mirror struts or mirror segment edges (Fig.~\ref{fig:interferogram}, \textit{Upper Left}) the holes share the same PSF envelope or ``primary beam'' shape.  This primary beam is modulated by the interferometric fringe pattern generated by the baselines between pairs of holes (See \citealt{Greenbaum2015}). The hexagonal hole shape leads to the overall hexagonal shape of the interferogram.}
The core of the WR~137 interferogram appears more extended than the PSF calibrator, which indicates the presence of resolved emission around WR~137.
\AS{This extended emission in the core of the WR~137 interferogram is a result of reduced fringe amplitudes caused by structure other than the bright central compact source.  The more-or-less isotropic nature of the extended emission results from additional structures in various directions from the dominant compact source. For example, if the target were instead a moderate contrast binary the image would look extended along the two components' separation vector.  In the case of WR~137, structure clearly extends more than a resolution element from the compact bright source. Note that fringe-phase information is difficult to see directly from the interferogram and requires additional tools for extraction.}

\subsection{Extracting Interferometric Observables with ImPlaneIA, SAMpip, and AMICAL}
\label{sec:extract}
The final stage of NIRISS AMI data processing (stage 3) includes the extraction of interferometric observables (squared visibilities and closure phases) from the calibrated exposures.
Details of this process are described in \citet{Sivaramakrishnan2023}.
At the time of this work, the third stage did not use the JWST science calibration pipeline for NIRISS AMI data.
The stage 3 processing tasks were conducted using three independent software packages: ImPlaneIA\footnote{\rev{\url{https://github.com/anand0xff/ImPlaneIA/commit/b6caf9db3b6976b3427b2a4ce5798e470f022c3d}}} \citep{Greenbaum2015,ImPlaneIA2018}, SAMpip \citep{SAMpip}, and AMICAL \citep{Soulain2020, soulain23ascl}.  Comparing the results across the three software tools was important for ensuring robust measurements given that observable extraction of AMI data has had limited testing with space-based observations. For each software tool, measured observables for the science target (WR~137) were extracted and then calibrated using the observables of the PSF calibration source (HD~228337).

There is an important detail to be noted about the data processing related to cosmic ray events. In the version of the JWST science pipeline used on this dataset, cosmic ray events are flagged at stage 1 (jump detection step); however, this did not apply to observations with NGROUPS less than 3. The F380M observations for this program, where NGROUPS = 2, notably do not meet this requirement and so data impacted by cosmic rays were passed on to the stage 3 processing steps. The presence of the cosmic rays can present issues with the image plane algorithms (ImPlaneIA and SAMpip).

All three software tools (ImPlaneIA, SAMpip, and AMICAL) extract interferometric observables from each integration and deliver the raw weighted mean and standard deviation in a final Optical Interferometry Flexible Image Transport System (OIFITS; 
\citealt{Pauls2005,Duvert2017}) file. SAMpip and ImPlaneIA create a model of the data by fitting the fringes directly in the image plane, and the interferometric observables are derived from the model's coefficients.  AMICAL instead computes the Fourier Transform of the interferogram and locates the position of the different spatial frequencies (and their corresponding visibility amplitudes and phases) using a matched filter template with the mask's geometry. Detailed information on the application of ImPlaneIA, SAMpip, and AMICAL to NIRISS AMI data is provided in Sec.~5.3 of \citet{Sivaramakrishnan2023}.

Lastly, calibrating out the instrumental transfer function of the squared visibilities is done by dividing the raw observables of the target by those of the calibrator star, and the calibration of the closure phases is done by subtracting the closure phases of the calibrator from those of the target.  
Figure \ref{fig:IOplots} shows the resulting calibrated interferometric observables for the science target with all three softwares using both the F380M and F480M observations. The calibrated squared visibilities derived from all three softwares are in close agreement, as are the calibrated closure phases from AMICAL and SAMpip. 
The cause for the discrepant closure phases derived from ImPlaneIA in both F380M and F480M observations is currently under investigation
\rev{by the ImPlaneIA development team. The discrepancy appears to be due to a difference in the ways ImPlaneIA and SAMpip compute the statistics of interferometric variables. ImPlaneIA determines errors in fringe phases, and separately, fringe amplitudes.  SAMpip calculates statistics using fringe complex visibilities before converting the complex errors into real number fringe phase and amplitude errors.  ImPlaneIA is currently being updated to use complex visibility averages, which will likely bring it into alignment with SAMpip's error estimation. The SAMpip and ImPlaneIA teams are working to quantify the effect of differences in the computed observables' statistics.
}

\begin{figure*}[t!]
    \centering
    \includegraphics[width=0.95\linewidth]{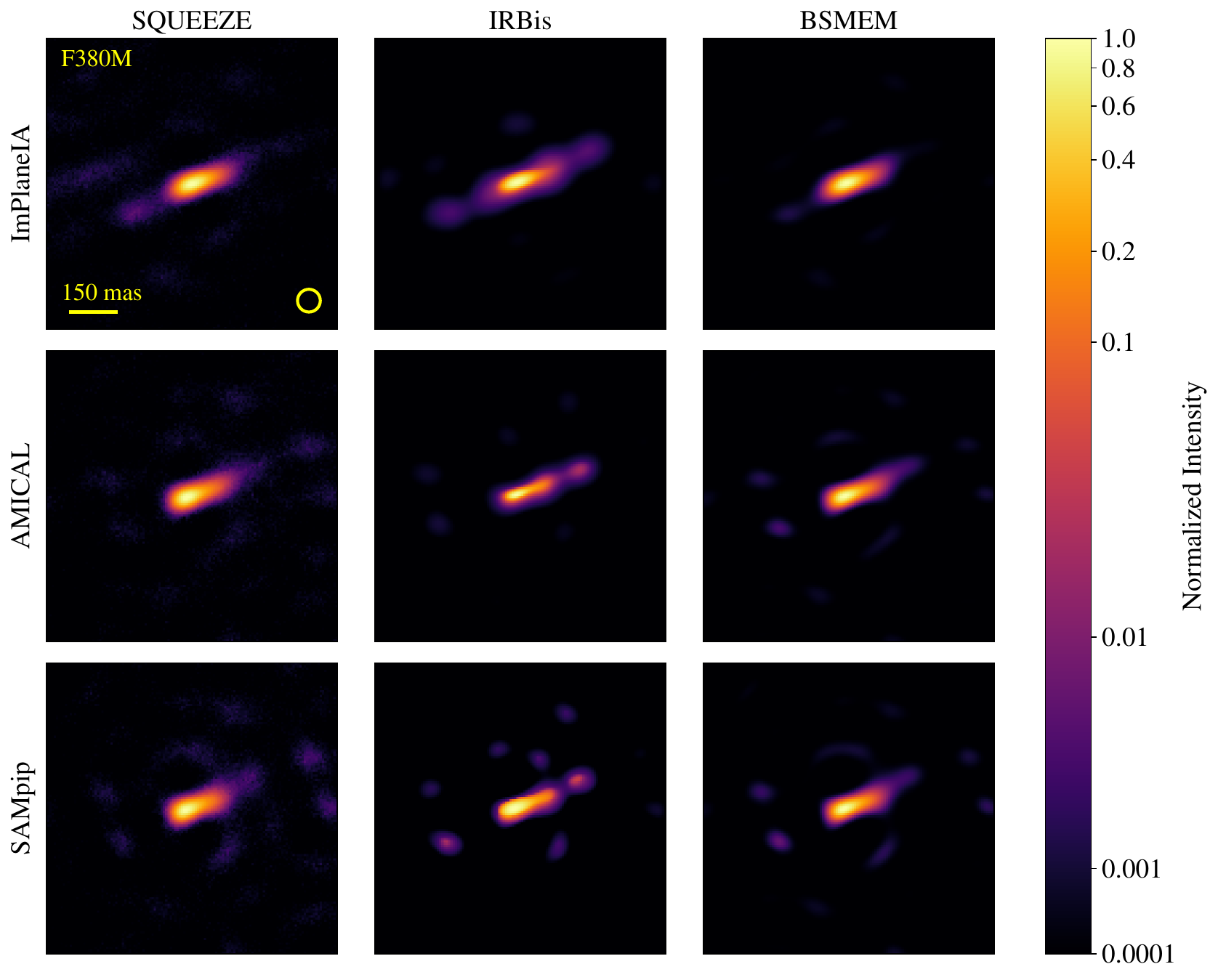}
    \caption{SQUEEZE, IRBis, and BSMEM image reconstructions of WR~137 based on the calibrated observables from the NIRISS AMI F380M observations extracted by ImPlaneIA, AMICAL, and SAMpip.  All images are shown with a logarithmic stretch normalized to the peak pixel intensity. The normalized intensity color bar is shown on the right. The angular resolution of the F380M images is indicated by the yellow circles in the upper left panel and corresponds to 60 mas. In all panels, north is up and east is to the left.}  
    \label{fig:ImReconF380M}
\end{figure*}

\begin{figure*}[t!]
    \centering
    \includegraphics[width=0.95\linewidth]{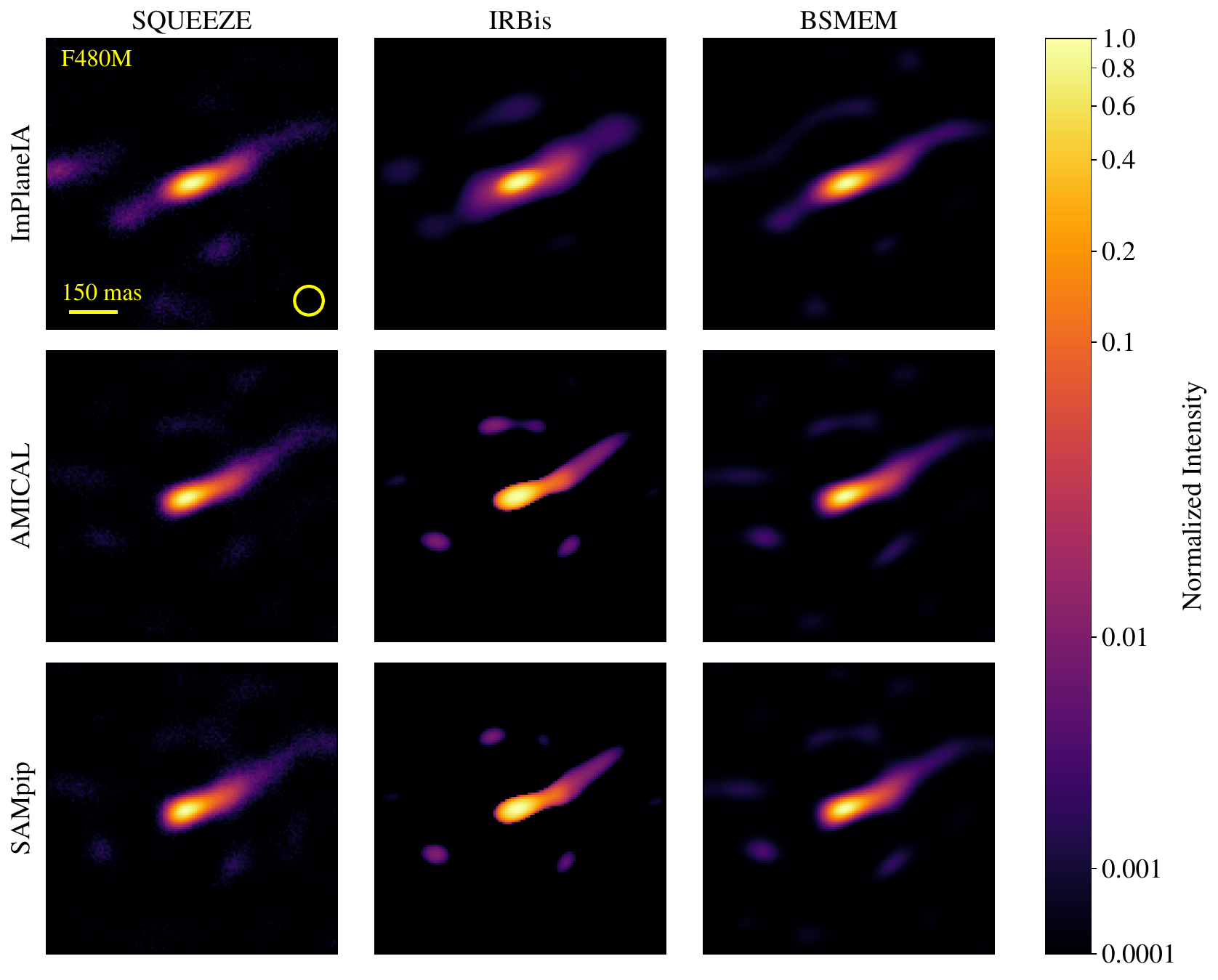}
    \caption{SQUEEZE, IRBis, and BSMEM image reconstructions of WR~137 based on the calibrated observables from the NIRISS AMI F480M observations extracted by ImPlaneIA, AMICAL, and SAMpip.  All images are shown with a logarithmic stretch normalized to the peak pixel intensity. The normalized intensity color bar is shown on the right. The angular resolution of the F480M images is indicated by the yellow circles in the upper left panel and corresponds to 80 mas.  In all panels, north is up and east is to the left.}  
    \label{fig:ImReconF480M}
\end{figure*}

\section{Results and Analysis}
\label{sec:RA}

\subsection{Image Reconstruction of WR~137's Circumstellar Environment}
\label{sec:imrec}

\begin{figure}[t!]
    \centering
    \includegraphics[width=0.98\linewidth]{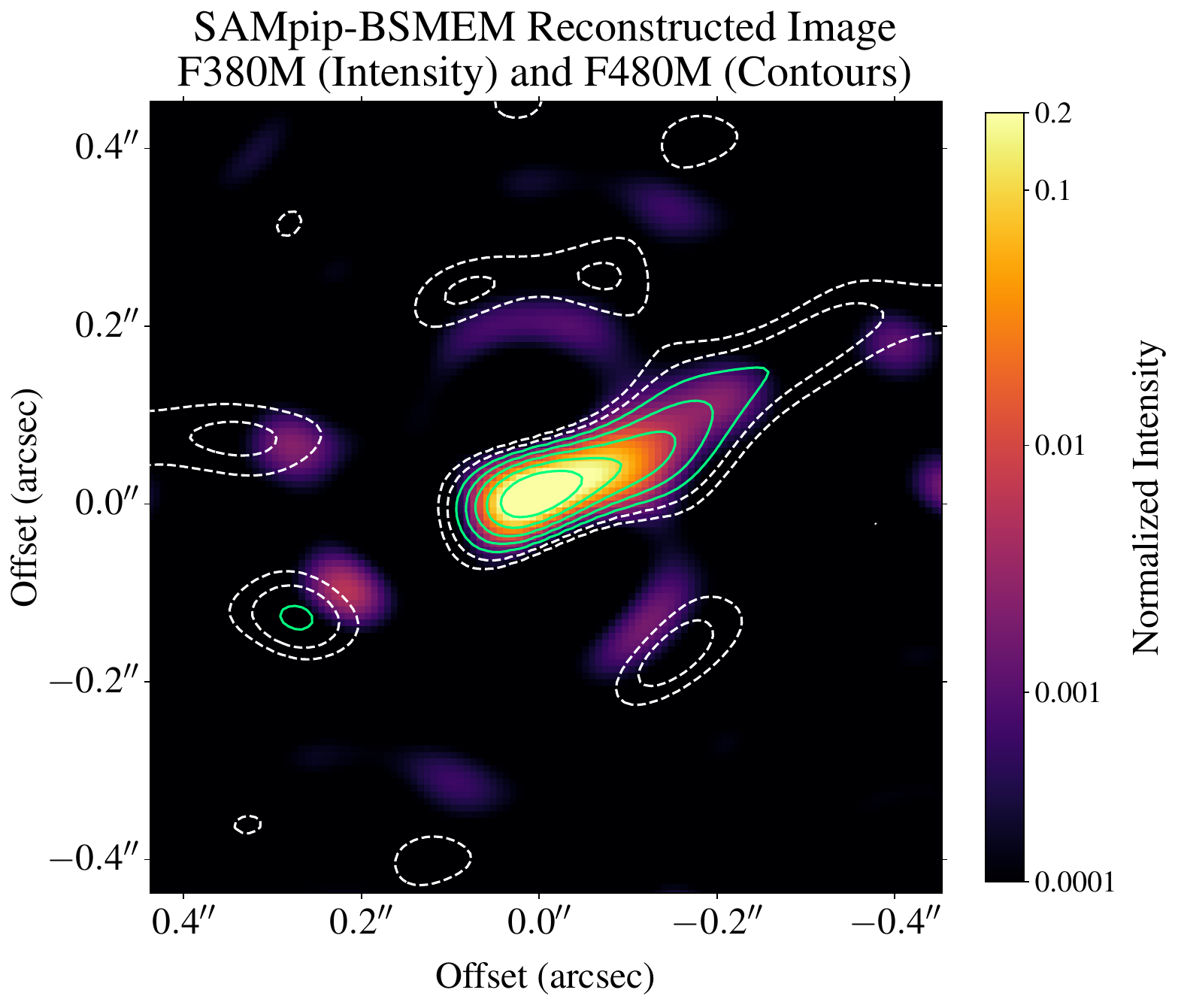}
    \caption{\rev{Reconstructed F380M (intensity) and F480M (contours) images of WR~137 using BSMEM on the observables extracted by SAMpip. The dashed, white contour levels correspond to normalized F480M intensities of $10^{-3.5}$ and $10^{-3.0}$, and the green, solid contours levels correspond to $10^{-2.5}$, $10^{-2.0}$, $10^{-1.5}$, $10^{-1}$, and $10^{-0.5}$. The $\lesssim10^{-2.5}$ features indicated by the dashed, white contours outside of the linear emission from WR~137 are likely not real given the wavelength dependence of their position.}}  
    \label{fig:noise}
\end{figure}

F380M and F480M images of WR~137 were reconstructed from the calibrated observables (Fig.~\ref{fig:IOplots}) using three different software tools: BSMEM \citep{Buscher1994,Lawson2004}, SQUEEZE \citep{Baron2010}, and IRBis \citep{Hofmann2014}.
Each of the image reconstruction software tools \rev{was} applied on the three sets of calibrated observables extracted by ImPlaneIA, AMICAL, and SAMpip. The reconstructed F380M and F480M images are presented in Figures~\ref{fig:ImReconF380M} \&~\ref{fig:ImReconF480M}, respectively.
This threefold approach provides independent methods of processing NIRISS AMI data and performing image reconstruction. The angular resolution achieved in these reconstructed images is $\sim0.5\,\lambda/D$, which corresponds to $\sim60$ mas and $\sim80$ mas for the F380M and F480M observations, respectively. \rev{An absolute photometric calibration was not performed on the reconstructed images due to the saturation of WR~137 in the direct NIRISS images (Sec.~\ref{sec:reduction}). However, future work is planned on strategies for performing an absolute photometric calibration of AMI observations targeting bright sources (e.g. utilizing the PSF reference star).  }

For each reconstructed image, similar image sizes and pixel scales corresponding to $128\times128$ pixels and 7.42 mas/pixel, respectively, were adopted. For the SQUEEZE reconstruction we employed entropy regularization and performed a simple grid search for a suitable hyper parameter value in the range from $\mu=10^{-3}$ to $\mu=10^3$ in logarithmic steps. \rev{For regularized minimization algorithms, the hyper parameter is a user-defined parameter that balances the weight between the likelihood and the prior probabilities when estimating the best-fit image.} The best-fit values for the hyper parameters were found with $\mu\leq10^{-1}$ with minimal variation in the range of $\mu=10^{-3}-10^{-1}$. Figures~\ref{fig:ImReconF380M} and~\ref{fig:ImReconF480M} present reconstructed images with hyper parameters of $\mu=10^{-2}$. The final SQUEEZE results presented are generated from an average of model chains which achieve a reduced chi-squared value $\chi_r^2\lesssim1.5$. 
The BSMEM reconstruction automatically selects the hyper parameter for the reconstruction. This software uses entropy as its regularization function, which tends to produce smooth brightness distributions. Furthermore, the entropy regularizer ensures that the pixel values are always positive. 
The $\chi_r^2$ values obtained are close to unity.

For IRBis, the image was reconstructed using the edge-preserving regularization function. An additional mask was also utilized to only use the flux within the mask radius $r_{mask}$. The reconstruction was then performed using a combination of $\mu$ (0.1 to $10^{-5}$) and $r_{mask}$ (550 to 640 mas). The prior image was a Gaussian model with a FWHM of 650 mas. The best reconstructed image was finally selected using the reconstruction quality parameters $q_\mathrm{rec}$ based on the $\chi_r^2$ and the residual ratio values $\rho\rho$ \citep[See][]{Hofmann2014}. The final IRBis results provided good agreement with the data for both F380M and F480M filters.
The IRBis results also presented a systematically better fit for the closure phases than the visibility amplitude.

Figures~\ref{fig:ImReconF380M} and~\ref{fig:ImReconF480M} demonstrate that all combinations of image reconstruction tool and observable extraction software produce similar reconstructed images of WR~137 in both wavelengths.
The dust emission extending to the NW from WR~137 appears quasi-linear with some slight curvature angled to the north. The extent of the linear emission is slightly shorter in the F380M images ($\sim200$ mas) compared to the F480M images ($\sim300$ mas). The detection of more extended emission in the F480M images is likely due to lower uncertainties in the F480M observables (Fig.~\ref{fig:IOplots}) \rev{and/or cooler dust temperatures at larger distances along the feature}.  
NIRISS AMI performance may be better in the F480M data than the F380M data because the 65 mas detector pixels better sample the F480M data than the shorter wavelength F380M data. 
The faint structure extending to the SE, which is most prominent in the IRBis reconstructions of the ImPlaneIA observables, is unlikely real given the absence of this feature in all other image \rev{reconstructions}. \rev{The appearance of the SE extension only in images reconstructed from the ImPlaneIA-calibrated observables is likely due to the slightly discrepant closure phase measurements (Fig.~\ref{fig:IOplots}). }

\rev{A comparison of real astrophysical features and likely artifacts from the image reconstruction is presented in Figure~\ref{fig:noise}, which shows the BSMEM-reconstructed image data from the F380M and F480M observables extracted by SAMpip. While the overlapping F380M and F480M emission of the bright, linear feature traces astrophysical emission from dust, the faint elliptical features that are displaced in radial position between the F380M and F480M images are most likely image reconstruction artifacts. }

The linear extended emission in the reconstructed images resembles previous observations of WR~137 obtained by ground-based mid-IR imaging \citep{Marchenko2007}. \rev{The orientation of this feature is also consistent with the alignment of dust clumps revealed by near-IR imaging with HST \citep{Marchenko1999}, which were obtained at a slightly later orbital phase ($\varphi\sim0-0.06$) than the NIRISS observations ($\varphi=0.9$). The near-IR clumps therefore likely trace dust density enhancements along a continuous, linear feature consistent with the feature revealed in the NIRISS observations. A shorter linear emission feature resolved in the near-IR HST images, however, extends in the opposite direction (southeast) of the NIRISS feature. If the origin of this feature is linked to colliding stellar winds (See Sec.~\ref{sec:cwmodel}) the opposite orientation is likely due to the different orbital configurations of the colliding-wind binary between the NIRISS and HST observations.}

\subsection{\rev{Colliding-wind Shock Opening Angle Analysis}}

\begin{figure*}[t!]
    \includegraphics[width=0.98\linewidth]{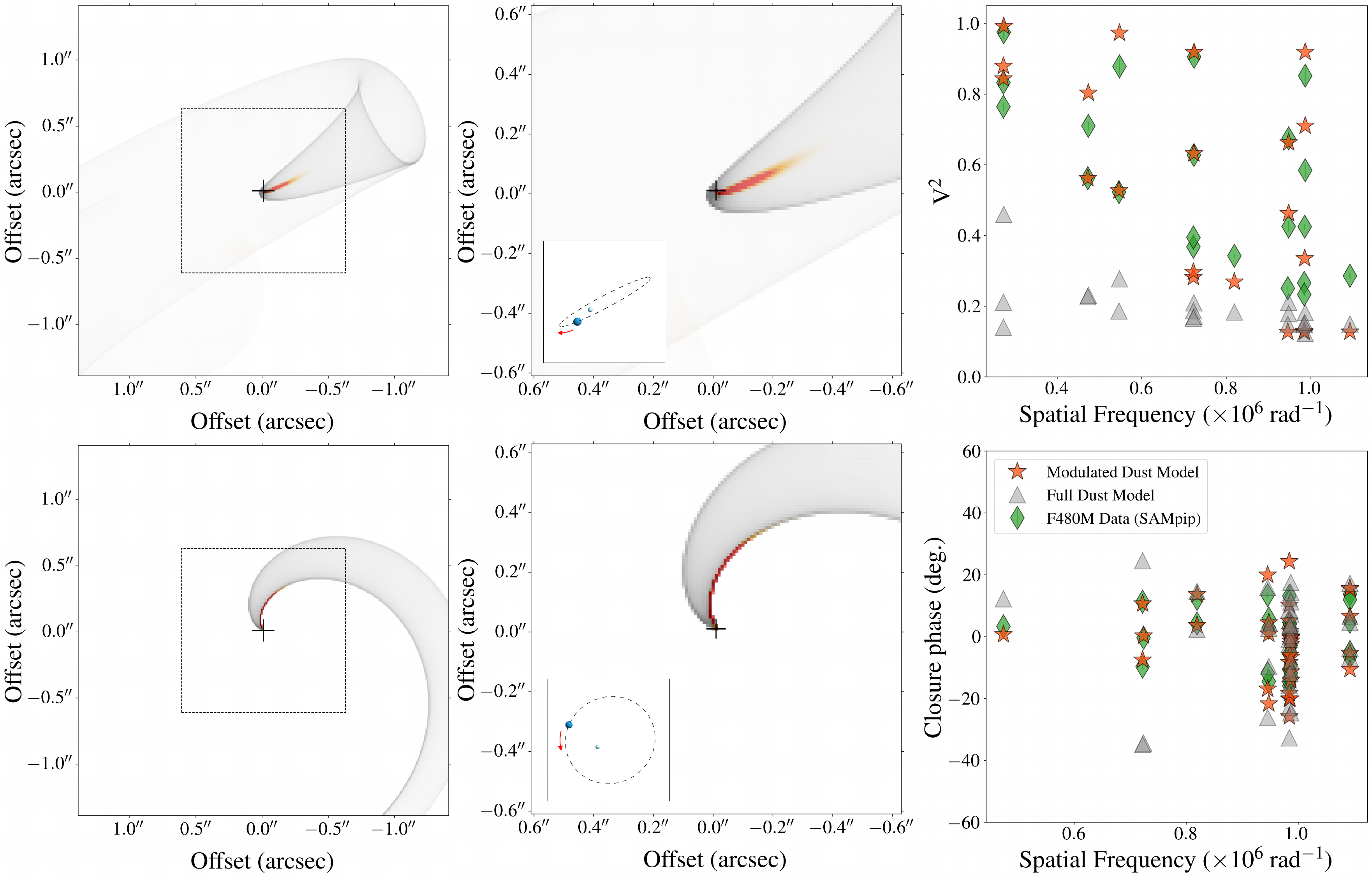}
    \caption{\textit{(Left and Center)} Observed and orbital plane projections of column densities from the full (Grey) and modulated (Red/Orange) geometric dust models of WR~137 using the parameters provided in Tab.~\ref{tab:orbit} at \rev{the} orbital phase \rev{of the} JWST/NIRISS observations \rev{($\varphi=0.90$)}. \rev{The inset of the zoomed images shows the orbit of the O9 star around the WC star at the phase of the JWST/NIRISS observations in the respective projections.} The crosses indicate the position of the unresolved binary system. \textit{(Right)} Simulated NIRISS AMI squared visibilities and closure phases of the full and modulated dust models compared to the real observables extracted from the F480M data with SAMpip. Observables from the modulated dust model show a closer agreement to the F480M data than the full dust model. Note that the error bars on the F480M observables are smaller than the symbols used. }
    \label{fig:Model}
\end{figure*}

\rev{The} linear morphology \rev{of WR~137's extended emission is different from} the extended dust emission \rev{morphologies resolved} around other dust-forming colliding-wind binaries \citep{Tuthill1999,Monnier2007,Lau2020WR112}. 
Such systems typically show structures consistent with dust formed in a hollow conical wind-collision region \rev{revolving} with the stars in their orbit and symmetrical about their line of centers. The conical shape of the ``shock cone'' assumes the collision of two isotropic winds, and its opening angle depends on the wind-momentum balance of the two stars (e.g.~\citealt{Usov1991, Canto1996}).
Assuming a purely hydrodynamic balance and spherically symmetric winds from the stars in WR~137, the half-opening angle of the thin shock cone ($\theta_h$), appropriate for radiative post-shock plasma, can be derived from the following expression (See \citealt{Canto1996,Tuthill2008}):

\begin{equation}
    \mathrm{tan}\, \theta_h - \theta_h = \frac{\eta\,\pi}{1-\eta},
\label{eq:theta}
\end{equation}

\noindent
where $\eta$ is the wind-momentum ratio of the companion star and the WR star:

\begin{equation}
    \eta = \frac{\dot{M}_\mathrm{OB} v_\mathrm{OB}}{\dot{M}_\mathrm{WR} v_\mathrm{WR}}.
\label{eq:eta}
\end{equation}
A shock-cone half-opening angle of $\theta_h=18.6^\circ$, where $\eta = 0.0038$, can be derived from the mass-loss rates and terminal wind velocities of the two stars in WR~137 inferred from Potsdam Wolf–Rayet (PoWR; \citealt{Hamann2004}) models presented by \citet{Richardson2016}: $\dot{M}_\mathrm{WR}=10^{-4.65}$ M$_\odot$ yr$^{-1}$, $v_\mathrm{WR}=1700$ km s$^{-1}$, $\dot{M}_\mathrm{OB}=10^{-7.1}$ M$_\odot$ yr$^{-1}$, and $v_\mathrm{OB}=1800$ km s$^{-1}$. 

If the dust emission around WR~137 traces the entire surface of its shock cone, an upper limit of $\sim8^\circ$ for the shock cone half-opening angle can be derived based on the extent of the linear emission and the angular resolution: $\sim300$ mas and 80 mas, respectively, for the F480M observations. 
The linear morphology of the extended emission from WR~137 therefore does not appear to be consistent with the predicted $18.6^\circ$ half-opening angle.
However, other factors can affect the dust morphology from colliding-wind binaries:
the OB/WR-wind momentum ratio $\eta$ may be overestimated, or the wind(s) may not be spherically symmetric thus leading to non-uniform dust formation across the surface of \rev{a possibly asymmetric} shock cone. 
Since it is difficult to produce such a linear morphology even with a much smaller $\eta$, it is unlikely the morphology is solely due to an overestimate of $\eta$. For example, even if $\eta$ were overestimated by a factor of 10, the full opening angle would be $\sim17^\circ$ and would be resolvable in the reconstructed images.
The impact of wind asymmetries is particularly important to consider given the likely presence of a decretion disk around the O9 companion \citep{St-Louis2020}. Variable dust formation across the surface of the shock \rev{interface} is therefore investigated in Sec.~\ref{sec:cwmodel} utilizing a geometric colliding-wind modeling tool.

\begin{deluxetable}{ll}
\tablecaption{Properties of WR~137 Geometric Dust Model}
\tablewidth{0pt}
\tablehead{\textbf{Adopted Properties} & }\
\startdata
Orbital Period ($P_\mathrm{Orb})$ & \rev{13.1} yr \\
\rev{Periastron Passage Date ($P_0$)} & \rev{2023.85} \\
Semi-major Axis ($a$) & \rev{8.56} mas\\
Eccentricity ($e$) & 0.315\\
Inclination ($i$) & \rev{97.2}$^\circ$\\
Argument of periapsis ($\omega$) & \rev{0.6}$^\circ$ \\
Longitude of the ascending node ($\Omega$) & \rev{117.91}$^\circ$ \\
Orbital phase ($\varphi$) & \rev{0.90} \\
Shock-cone half-opening angle ($\theta_h$) & 18.6$^\circ$ \\
Distance to WR~137 (d) & 1941 pc\\
Dust expansion velocity ($v_\mathrm{exp}$) & 1700 km s$^{-1}$\\
Azimuthal dust modulation centroid ($\mu_\mathrm{Az}$) & 180$^\circ$ \\
\hline
\textbf{Free Parameters}\\
\hline
Azimuthal dust modulation width 
 ($\sigma_\mathrm{Az}$) & \rev{6}$^\circ$\\
Orbital dust modulation centroid ($\mu_\mathrm{Orb}$) & \rev{265}$^\circ$  \\
Orbital dust modulation width ($\sigma_\mathrm{Orb}$) & \rev{13}$^\circ$ \\
\enddata
\tablecomments{\rev{The orbital period was adopted from \citet{Lefevre2005} and Richardson et al.~(in prep)}, and the other orbital parameters (\rev{$P_0$}, $a$, $e$, $i$, $\omega$, and $\Omega$) were adopted from recent CHARA observations of WR~137 (Richardson et al.,~in prep). The dust expansion velocity is assumed to be consistent with the wind velocity of the WC star provided by \citet{Richardson2016}, and the half-opening angle is derived from the wind momentum balance between the WC and companion star (Eq.~\ref{eq:theta}, \citealt{Richardson2016}). The adopted orbital phase is consistent with the expected orbital phase at the time of the observations, 15 July 2022 (Tab.~\ref{tab:Obs}). The azimuthal dust modulation centroid was assumed to be $\mu_\mathrm{Az}=0^\circ$. The remaining three orbital and azimuthal dust modulation parameters ($\mu_\mathrm{Orb}$, $\sigma_\mathrm{Orb}$, and $\sigma_\mathrm{Az}$) were the only free parameters in the model and were adjusted until a satisfactory agreement with the calibrated observables from the F480M observations was achieved (Fig.~\ref{fig:Model}, \textit{Right}).}
\label{tab:orbit}
\end{deluxetable}

\subsection{Interpreting WR~137's Extended IR Emission with Geometric Colliding-wind Models}
\label{sec:cwmodel}

Geometric modeling of dust production from colliding-wind binaries provides an important tool to interpret the spatially resolved dust emission around dust-forming WC binaries \citep{Williams2009,Callingham2019,Lau2022,Han2022}.
Such models, which output a map of dust column density, can assess whether the extended emission from WR~137 is consistent with dust production from a colliding-wind binary. 
The inputs required for the geometric dust modeling of WR~137 are the distance ($d$), shock-cone half-opening angle ($\theta_h$), dust expansion velocity ($v_\mathrm{exp}$), and orbital parameters of the binary systems which include the orbital period ($P_\mathrm{Orb}$), \rev{time of periastron passage ($P_0$),} inclination ($i$), semi-major axis ($a$), eccentricity ($e$), argument of periapsis ($\omega$), the longitude of the ascending node ($\Omega$), and an orbital phase at a given point in time ($\varphi$). 

Orbital parameters were adopted from recent CHARA observations with MIRC+CLIMBX that resolved the binary components in WR~137 (Richardson et al., in prep) and also incorporated previous CHARA observations by \citet{Richardson2016}. The orbital parameters are provided in Table~\ref{tab:orbit} and assume that the O9 companion star is brighter than the WC7 star in the near-IR CHARA observations. If instead the WC7 star is assumed to be brighter than the O9 companion, there would simply be $180^\circ$ added to the argument of periapsis ($\omega$). The value of \rev{$\omega=0.6^\circ$}, which assumes the O9 star is brighter than the WC7 star, is notably in closer agreement with the value derived independently from radial velocity observations by \citet{Lefevre2005} ($\omega=326\pm15^\circ$) than \rev{when} the WC7 star is assumed to be the brighter near-IR component ($\omega=180.6^\circ$).

The effects of non-uniform dust formation can be investigated with the geometric models by modulating dust formation across the surface of the shock cone in the orbital and azimuthal\footnote{i.e.~perpendicular to the WR--O-star axis.} directions (See Fig.~3 of \citealt{Han2022}). Orbitally modulated and azimuthally asymmetric dust production \rev{was notably} inferred \rev{in} the colliding-wind WR binary WR~140 \citep{Williams2009,Han2022}.
As \rev{in} \citet{Han2022}, dust production variability in the orbital and azimuthal directions are each modeled by a Gaussian distribution of dust density as a function of true anomaly and azimuthal angle, respectively. 
The parameters of modulated dust production are the centroids ($\mu$) and widths ($\sigma$) of the Gaussian functions in the orbital and azimuthal directions.
\rev{The model images were calculated at a spatial scale of 10 mas/pixel.}

An azimuthal centroid of $\mu_\mathrm{Az}=180^\circ$ was adopted, which corresponds to enhanced dust formation along the orbital plane in the ``trailing arm'' of the shock cone (Fig.~\ref{fig:Model}, \textit{Left} and \textit{Center}). 
This assumption is based on dust formation in WR~140, where dust emission is enhanced in the trailing arm \citep{Williams2009,Han2022}.
It is important to note that an asymmetry in dust production between the leading and trailing arms is indeed predicted based on 3D hydrodynamic modeling \citep{Hendrix2016}, however, such models of the colliding-wind binary WR~98a by \citet{Hendrix2016} instead predict dust enhancement in the leading arm.
Dust enhancement along the leading arm of WR~137's shock cone (i.e.,~$\mu_\mathrm{Az}=0^\circ$) cannot be conclusively ruled out due to degeneracies in geometric model parameters. 
Multi-epoch observations resolving the changing dust morphology around WR~137 will be important for resolving these degeneracies and identifying the region of enhanced dust formation.

In order to compare with the NIRISS AMI observations of WR~137, the dust column density maps output from the geometric modeling tool were converted to intensity assuming isothermal and optically-thin dust emission. Given the higher signal-to-noise ratio of the F480M observations, the column density maps were converted to intensity at 4.8 $\mu$m. Emission from the unresolved binary, which is not incorporated in the geometric models, is included by modifying the central pixel of the dust map output from the geometric modeling tool. 
Since dust emission is coincident along the line-of-sight with stellar emission from the binary, it is not possible to distinguish the stellar and dust emission from the NIRISS observations. However, archival mid-IR light curves that sampled WR~137 throughout its orbit can be used to estimate the emission from the central binary and circumstellar dust \citep{Williams2001}. The stellar emission at 4.8 $\mu$m, $F^*_{4.8\mu\mathrm{m}} = 0.8$ Jy, can be derived from the quiescent $L'$-band ($\lambda=3.8$ $\mu$m) flux density, $F^*_{3.8\mu\mathrm{m}}=1.7$ Jy, and the $\lambda F^*_\lambda\propto \lambda^{-1.86}$ power law that characterizes the IR spectrum of its quiescent emission \citep{Williams2001}. The total 4.8 $\mu$m emission, $F^\mathrm{Tot}_{4.8\mu\mathrm{m}} = 2.8$ Jy is estimated from $M'$-band ($\lambda=4.7$ $\mu$m) photometry of WR~137 obtained by \citet{Williams2001} in March 1996, which corresponds to a similar orbital phase of WR~137 as that of the JWST NIRISS observations ($\varphi\sim0.9$). 
The dust column density map and the stellar component were simply scaled to $70\%$ and $30\%$ of $F^\mathrm{Tot}_{4.8\mu\mathrm{m}}$, respectively, to produce a 4.8 $\mu$m intensity model of WR~137.

\begin{figure}[t!]
    \includegraphics[width=1.0\linewidth]{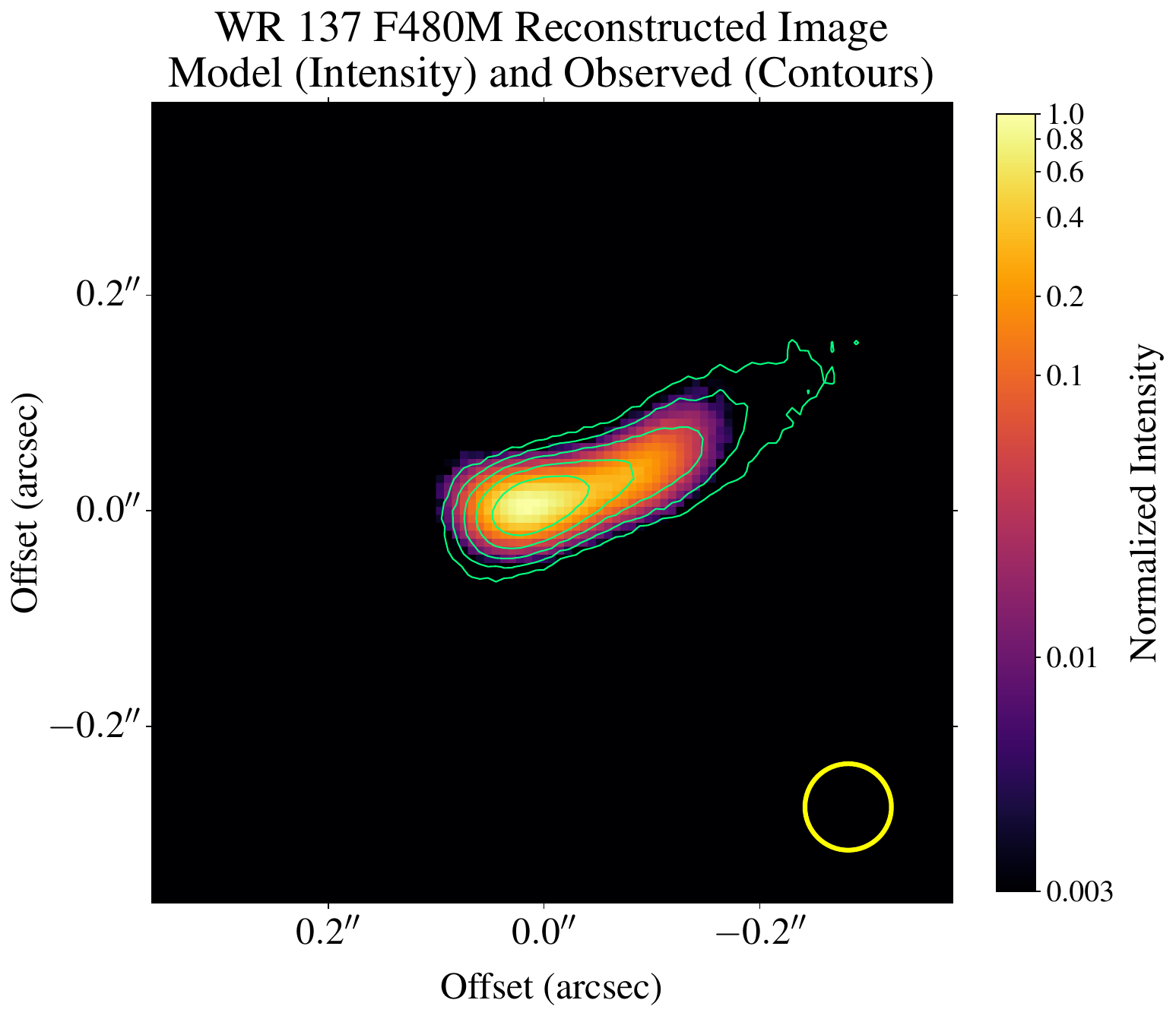}
    \caption{SQUEEZE image reconstruction of the modulated dust model with contours of SQUEEZE-reconstructed image from SAMpip F480M observables. The contour levels correspond to normalized intensities of \rev{$10^{-2.5}$}, $10^{-2}$, $10^{-1.5}$, $10^{-1}$, and $10^{-0.5}$. The yellow circle corresponds to the angular resolution of the F480M observations ($\sim80$ mas)}
    \label{fig:mod_imrec}
\end{figure}

Synthetic interferometric observables were extracted from the model 4.8 $\mu$m intensity maps of WR~137 to \rev{compare} with the NIRISS F480M observables. Specifically, the F480M observables obtained using SAMpip (Fig.~\ref{fig:IOplots}) were used to compare the synthetic observables. The synthetic observables were extracted by computing the Discrete Fourier Transform of the model image using the spatial frequencies sampled with NIRISS AMI. From the Fourier amplitudes and phases, the squared visibilities and closure phases \rev{were} constructed using the baseline information of the non-redundant mask. Uncertainties on the synthetic squared visibilities and closure phases were adopted from the SAMpip F480M observables. Note that the extraction of synthetic F480M observables assumes a single wavelength ($\lambda=4.8$ $\mu$m) for the dust emission from the model.

With the azimuthal centroid fixed at $\mu_\mathrm{Az}=180^\circ$, the remaining three free dust modulation parameters ($\sigma_\mathrm{Az}$, $\mu_\mathrm{Orb}$, and $\sigma_\mathrm{Orb}$) were adjusted to find a satisfactory agreement with the SAMpip F480M closure phases and squared visibilities.
Figure~\ref{fig:Model} (\textit{Left} and \textit{Center}) presents the dust column density maps from the geometric modelling of WR~137 with and without incorporating modulated dust production parameters from Table~\ref{tab:orbit}. The interferometric observables of the geometric dust models and the F480M observations presented in Fig.~\ref{fig:Model} (\textit{Right}) show that the observables from the model without modulated dust production disagree with that of the observations, as expected. Fig.~\ref{fig:Model} (\textit{Right}) demonstrates the general agreement between the observations and the modulated dust production model where dust production is confined to the orbital plane and is enhanced as the system approaches periapsis.

A reconstructed image using SQUEEZE with a similar configuration used for the F480M image reconstruction (See Sec.~\ref{sec:imrec}; Fig.~\ref{fig:ImReconF480M}) was generated from the synthetic observables of the modulated geometric dust model. Figure~\ref{fig:mod_imrec} presents the reconstructed image of the modulated dust model overlaid with the contours of the SQUEEZE-reconstructed image from the SAMpip F480M calibrated observables (See Fig.~\ref{fig:ImReconF480M}). 
The morphology and intensity profile of the extended dust emission from the modulated dust model show a close agreement to the F480M observations. This agreement suggests that extended dust emission from WR~137 can be explained by dust-production via the colliding-wind mechanism with modulated dust formation rates along the surface of the shock cone.

\section{Discussion\rev{: Dust-formation Enabled by Enhanced Equatorial Mass-loss from the O9 Companion?}}
\label{sec:discussion}

\rev{Observations of persistent double-peaked emission line profiles from the O9 companion star in WR~137 and its polarization signature indicate that the O9 star has a decretion disk that may be linked to the star's rapid rotation \citep{Harries2000,Richardson2016,St-Louis2020}.}
The enhancement of dust production along WR~137's orbital plane (Fig.~\ref{fig:Model}) may \rev{therefore} be influenced by \rev{anisotropic} wind densities and velocities \rev{that differ} from those of a non-rotating O star. \rev{The} polarimetric observations of WR~137 \rev{notably indicate} the disk \rev{is} aligned with the orbital plane of the system \citep{Harries2000,St-Louis2020}. It is therefore plausible that an anisotropic, equatorially-enhanced wind from the O9 companion presents more favorable conditions for dust-formation via the colliding-wind mechanism along the orbital plane due to enhanced \rev{densities}. 
The potential influence of anisotropic winds on colliding-wind dust formation \rev{has} been observed \rev{in} another WR binary, Apep \citep{Callingham2019,Han2020}.
\rev{Recent spectroscopic mid-IR observations of WR~137 with the Stratospheric Observatory for Infrared Astronomy (SOFIA) also suggest that the interaction between the WC and O9 winds is important for triggering dust formation \citep{Peatt2023}.}

In order to investigate dust-forming conditions in the colliding-wind shock, we utilize the dimensionless parameter, $\Gamma$, defined by \citet{Usov1991} to characterize the radiative cooling of gas in the shock layer:

\begin{multline}
    \Gamma \simeq 0.8 \left(\frac{\dot{M}_\mathrm{WR}}{10^{-5}\,\mathrm{M}_\odot\,\mathrm{yr}^{-1}}\right) \left(\frac{D}{0.67\,\mathrm{au}}\right)^{-1}  \left(\frac{v_\mathrm{WR}}{10^{3}\,\mathrm{km}\,\mathrm{s}^{-1}}\right)^{-3}\\  \times (1 + \eta^{1/2})\,\eta^{1/2},
    \label{eq:gamma}
\end{multline}

\noindent
where $D$ is the \rev{instantaneous} separation between the WR and the companion star, and $\eta$ is the  momentum ratio of the companion star \rev{wind} over \rev{that of} the WR star (Eq.~\ref{eq:eta}). Higher values of $\Gamma$ indicate increased cooling of hot gas in the shock. \rev{The} radiative cooling efficiency is an important factor in dust formation \rev{since} the hot $\sim10^{7}-10^{8}$ K gas must cool to $\sim1000$ K in order to form dust. 
We note that $\Gamma$ does not account for clumping, which likely plays a significant role in colliding-wind dust formation \citep{Eatson2022}. However, as a simplified investigation of dust-formation in colliding winds, we assume that
\rev{WR~137 forms dust for a range of $\Gamma$ values similar to the values of $\Gamma$ derived from dust formation episodes of WR~140. WR~140 is a well-studied, dust forming colliding wind binary, which forms dust at a phase interval $\varphi_\mathrm{dust} = \pm 0.04$ around periastron passage \citep{Han2022}.}
Based on Eq.~\ref{eq:gamma} and adopting stellar and orbital properties of WR~140 from previous literature, we calculate $\Gamma$ for WR~140, $\Gamma_\mathrm{WR140}$, and normalize by $\Gamma$ at the separation where WR~140's dust formation begins/ends, $D_\mathrm{dust}=8.3$ au\footnote{ \rev{Assuming the Gaia distance to WR~140 of 1.64 kpc \citep{Thomas2021}}}:

\begin{equation}
    \frac{\Gamma_\mathrm{WR140}(D)}{\Gamma_\mathrm{WR140}(D_\mathrm{dust})}=1.0 \left(\frac{D}{8.3\,\mathrm{au}}\right)^{-1}.
    \label{eq:gammaWR140}
\end{equation}

\noindent
We adopt a WR mass-loss rate, terminal wind velocity, and OB/WR wind momentum ratio of $\dot{M}_\mathrm{WR}=2\times10^{-5}$ \citep{Sugawara2015}, $v_\mathrm{WR}=2860$ km s$^{-1}$ \citep{Williams1989}, and $\eta = 0.043$ \citep{Han2022} for WR~140, respectively.

It is important to emphasize that utilizing $\Gamma$ as a dust formation diagnostic does not capture all of the complex physics of dust condensation in colliding winds. \rev{For example,} although $\Gamma_\mathrm{WR140}$ increases as WR~140 approaches periastron ($\varphi = 0$; $D = 1.5$ au), observations demonstrate that dust formation at separation distances $D<D_\mathrm{dust}$ appears to \textit{decrease} relative to $D = D_\mathrm{dust}$ \citep{Williams2009,Han2022}.
\rev{When stars are close,} ``sudden radiative braking" of the WR wind due to deceleration \rev{by the photospheric UV emission of the O star} \citep{Gayley1997} likely \rev{plays} an important role in mitigating dust production at close separation distances. However, an investigation of these effects is beyond the scope of this work.

Using the orbital parameters of WR~137 from Table~\ref{tab:orbit} and the mass-loss rates and terminal wind velocities from \citet{Richardson2016}, we can calculate $\Gamma_\mathrm{WR137}$ normalized by $\Gamma_\mathrm{WR140}(D_\mathrm{dust})$ as follows:

\begin{equation}
    \frac{\Gamma_\mathrm{WR137}(D)}{\Gamma_\mathrm{WR140}(D_\mathrm{dust})}=0.82 \left(\frac{D}{13.2\,\mathrm{au}}\right)^{-1},
    \label{eq:alphaWR137}
\end{equation}

\noindent
where \rev{$D=13.2$} au corresponds to the binary separation in WR~137 at the time of the JWST observations \rev{at an orbital phase of} \rev{$\varphi=0.90$}.

\begin{figure}[t!]
    \includegraphics[width=0.98\linewidth]{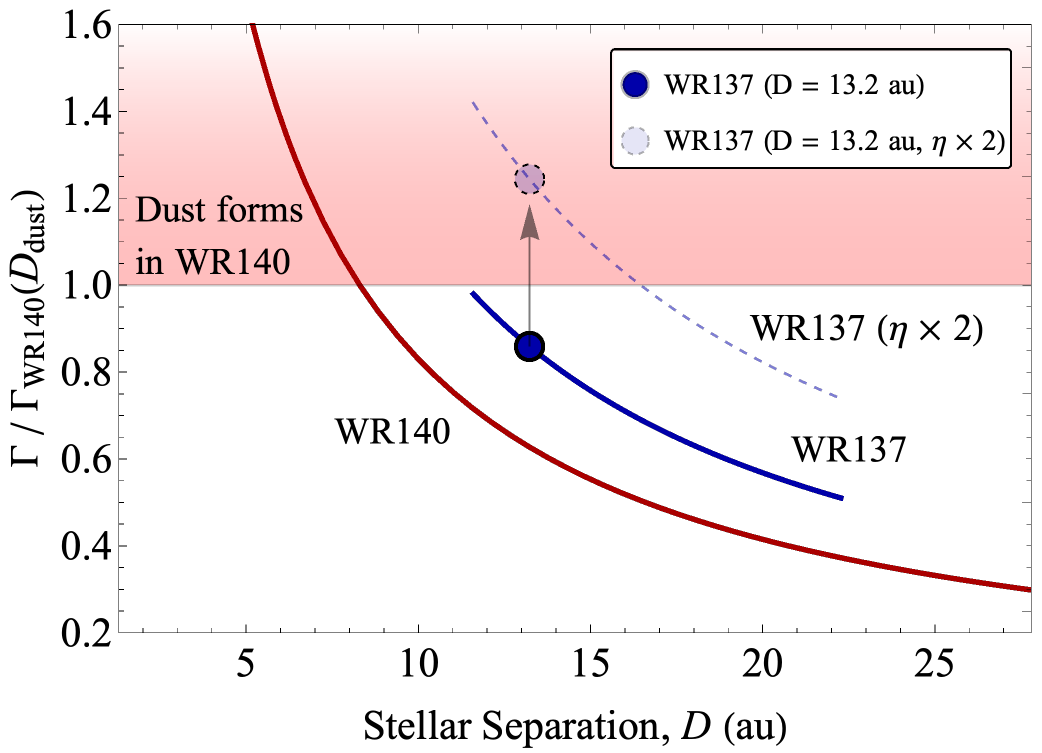}
    \caption{Estimated colliding-wind dust-formation thresholds for WR~137 and WR~140 as a function of binary separation, $D$, using the dimensionless radiative cooling efficiency parameter, $\Gamma$, as a diagnostic. $\Gamma$ is normalized by the value when dust formation begins/ends in WR~140, $\Gamma_\mathrm{WR140}(D_\mathrm{dust} = 8.3\, \mathrm{au})$. The shaded region above $\Gamma/ \Gamma_\mathrm{WR140}(D_\mathrm{dust})=1$ corresponds to where dust formation is expected based on observations of WR~140 and its orbital parameters; however, note that other factors such as sudden radiative braking may inhibit dust production at close separation distance where $\Gamma/ \Gamma_\mathrm{WR140}(D_\mathrm{dust})\gg1$. WR~137 may not be expected to form dust unless the wind momentum ratio of its O-star companion to the WC star, $\eta$ (Eq.~\ref{eq:eta}), is enhanced by a factor of $\sim2$ due to a rapidly rotating O-star companion (e.g.~\citealt{Maeder2000})} 
    \label{fig:Gamma}
\end{figure}

A comparison of $\Gamma/ \Gamma_\mathrm{WR140}(D_\mathrm{dust})$ for WR~140 and WR~137 throughout the \rev{stellar separation} of their respective orbits is shown in Figure~\ref{fig:Gamma}. Despite the clear evidence for dust production at the time of the JWST observations (Fig.~\ref{fig:ImReconF380M}~\&~\ref{fig:ImReconF480M}), Fig.~\ref{fig:Gamma} suggests that throughout its entire orbit, the radiative cooling of shocked gas in WR~137 \rev{is not strong enough to allow} for dust production with the adopted stellar and orbital properties. 
However, the adopted mass-loss rates and wind velocities of WR~137 \citep{Richardson2016} were \rev{derived} assuming spherical symmetry \rev{and thus do not} not account for enhanced equatorial mass-loss from the O-star companion. 
Interestingly, \citet{Maeder2000} predict that the equatorial mass-loss rate is enhanced by a factor of $\sim2$ for a rapidly rotating O-star (See Table 1 of \citealt{Maeder2000}).
If the mass-loss rate and/or terminal wind velocity of the companion O-star in WR~137 were enhanced by a factor of $2$, the radiative cooling in the colliding-winds would satisfy the dust-formation threshold at the \rev{stellar separation at the time} of the JWST observations (Fig.~\ref{fig:Gamma}).

We note that the isotropic O-star mass-loss rates from \citet{Richardson2016} have large uncertainties ($\mathrm{log}\,\dot{M}=-7.1^{+1.0}_{-0.3}$) \rev{so} that the upper \rev{limit} would \rev{allow dust to form.} However, a larger isotropic mass-loss rate from the O star would also lead to a larger value of $\eta$ and a wider shock-cone opening angle (See Eq.~\ref{eq:theta}), which is not consistent with the linear morphology of the dust emission in the NIRISS observations (Fig.~\ref{fig:ImReconF380M}~\&~\ref{fig:ImReconF480M}). 
We therefore argue that colliding-wind dust production in WR~137 is enabled by enhanced equatorial mass-loss from the O9 companion.
Rapid stellar rotation may therefore promote dust production in colliding-wind binaries, and circumstellar dust emission confined to the orbital and/or equatorial plane of the rapidly rotating star likely \rev{indicates the impact of} this effect.

\rev{It is important to address the following caveats on the interpretations and discussion. The analysis of dust-forming conditions using the $\Gamma$ parameter notably does not consider a possible reduction in the mass-loss rate or wind velocity that would impact the wind momentum ratio $\eta$. Asymmetric winds would also alter the morphology of the colliding-wind shock cone that would deviate from the geometric model presented in Fig.~\ref{fig:Model}. Additionally, it is unclear if rapid stellar rotation can result in enhanced mass-loss confined to such a narrow angular region ($\sim6^\circ$; Tab.~\ref{tab:orbit}) around the equator.}
Future work with hydrodynamical simulations (e.g.~\citealt{Eatson2022}) that can capture the complex physics of colliding-wind dust formation with asymmetric wind(s) \rev{will be important for investigating if non-spherical mass loss can indeed promote dust production in colliding-wind binaries}

\section{Summary}

In this paper, we presented JWST observations of the periodic dust-forming colliding-wind binary system WR~137 using the AMI mode of NIRISS with the F380M and F480M filters. The observations were taken as part of the WR DustERS program (ERS 1349) and provide some of the first science results using the NIRISS AMI observing mode on JWST. Notably, the WR~137 results demonstrate that NIRISS AMI is uniquely suited for observations of faint and extended mid-IR emission around a bright central core at angular scales of $\lesssim400$ mas. The NIRISS AMI observations of WR~137 were obtained with two different dither strategies (See Sec.~\ref{sec:Obs}). \rev{The} analysis in this work was performed on the `stare' (non-dithered) observations.

\rev{We extracted interferometric observables} from the interferogram pattern from the WR~137 observations and calibrated against a PSF-calibrator star (HD~228337) using three different software packages: ImPlaneIA, SAMpip, and AMICAL.  The calibrated squared visibilities of the WR~137 observations were consistent across all three software \rev{packages} (Fig.~\ref{fig:IOplots}). The calibrated closure phases from AMICAL and SAMpip were consistent as well, \rev{while closure phases derived from ImPlaneIA were slightly discrepant.} The cause for the discrepant closure phases is currently under investigation. The F480M observations yielded observables with the smallest uncertainties, which is likely due to the larger number of NGROUPS in each integration than in the F380M integrations (Tab.~\ref{tab:Obs}). 

Images of WR~137 were reconstructed from the interferometric observables using three different software tools: BSMEM, SQUEEZE, and IRBis. Each tool utilized the three sets of observables extracted from the \rev{three} software \rev{packages}. 
The reconstructed F380M and F480M images presented a nearly identical picture of WR~137 with a bright central core and with a $\sim200-300$ mas quasi-linear filament extending toward the northwest (Fig.~\ref{fig:ImReconF380M}~\&~\ref{fig:ImReconF480M}). The similarity of the images reconstructed with \rev{these three} independent tools demonstrates the robustness of capturing WR~137's morphology from the NIRISS AMI observations.

\rev{The expected half-opening angle of the shock cone in WR~137 is $\theta_h=18.6^\circ$ (See Sec.~\ref{sec:cwmodel}), but a shock cone that wide is not consistent with the linear morphology of the resolved dust emission.} \rev{We used the geometric colliding-wind dust modelling tool from \citet{Han2022}} to interpret the linear dust morphology extending from WR~137 \rev{using} orbital parameters from recent CHARA observations by Richardson et al. (in prep; Tab.~\ref{tab:orbit}). In order to reproduce the linear appearance of the extended dust emission, variable dust production rates across the surface of the colliding-wind shock cone in the orbital and azimuthal directions \rev{were needed}. A geometric model with dust formation confined to the orbital plane and enhanced as the system approaches periapsis provided a closer agreement to the interferometric observables from the F480M observations than the full model without any dust-production variability (Fig.~\ref{fig:Model}). An image reconstructed by SQUEEZE from the modulated dust-model observables \rev{showed} a close resemblance to the reconstructed image from the F480M observations (Fig.~\ref{fig:mod_imrec}).

We discussed the possible \rev{effect} of \rev{enhanced equatorial mass-loss from the} O9 companion star in WR~137 \citep{Richardson2016,St-Louis2020} \rev{to explain the} linear morphology of the observed dust production. We used the \rev{analytical} colliding-wind dust production \rev{framework} by \cite{Usov1991} and the well-studied orbital and stellar properties of WR~140 as a reference to investigate dust-formation in WR~137. As a diagnostic for dust formation in colliding-winds, we used estimated values of the parameter $\Gamma$ which characterizes the radiative cooling of gas in the colliding-wind shock layer \citep{Usov1991}. We found that WR~137 should not be capable of forming dust given its orbital and stellar properties (Fig.~\ref{fig:Gamma}). However, \rev{if the equatorial mass loss from the} rapidly rotating O9 companion star were enhanced by a factor of $\sim2$ \citep{Maeder2000}, \rev{this} would be sufficient to enable dust formation in WR~137 (Fig.~\ref{fig:Gamma}). We therefore conclude that equatorially enhanced mass-loss from the rapidly rotating O9 companion star \rev{may be} responsible for the dust formed along the orbital plane of WR~137, which is aligned with the rotation axis of the O9 star \citep{St-Louis2020}. 

JWST observations of WR~137 with NIRISS AMI provided us with the imaging contrast and sensitivity at mid-IR wavelengths to perform a morphological analysis of dust-forming conditions in colliding winds. 
Our results \rev{present a first look at the capabilities of NIRISS AMI that indicate its potential for investigating} a wide range of astrophysical environments having bright central cores and faint and close-in extended emission, including active galactic nuclei, evolved stars, and young stellar objects. 
Notably, the dynamic range of NIRISS AMI observations is expected to improve as technical studies progress for optimizing the analysis and calibration of these datasets, which will allow for probing even fainter features \rev{(See \citealt{Ray2023,Sallum2023})}.
Sub-pixel dithering \citep{Fruchter2002} and improved calibration of second-order systematics such as charge migration may also bring F380M and F430M data up to the quality seen in F480M (See Sec.~\ref{sec:imrec}).  
In this work, our aim was not only to investigate colliding-wind dust production, but also to provide a first look at science with the NIRISS AMI observing mode on JWST and to set the stage for the future of space-based aperture-masking observations.

\begin{acknowledgments}
RML would like to acknowledge the members of the entire WR DustERS team for their valuable discussions and contributions to this work. 
We thank Amaya Moro-Martin, William Januszewki, Neill Reid, Margaret Meixner, and Bonnie Meinke for their support of the planning and execution of our DD-ERS program. We would also like to acknowledge the NIRISS instrument and MIRaGe teams for their support of our observation preparation and data analysis plans.
We also thank Tomer Shenar for the correspondence on the stellar wind models of WR~137.
\rev{We would also like to acknowledge the anonymous referee for their insightful feedback that has improved the quality and clarity of this work.}
AFJM is grateful to NSERC (Canada) for financial aid.
NDR is grateful for support from the Cottrell Scholar Award \#CS-CSA-2023-143 sponsored by the Research Corporation for Science Advancement. J.S.-B. acknowledges the support received from the UNAM PAPIIT project IA 105023; and from the CONAHCyT “Ciencia de Frontera” project CF-2019/263975.

\end{acknowledgments}


This work is based on observations made with the NASA/ESA/CSA James Webb Space Telescope. The data were obtained from the Mikulski Archive for Space Telescopes at the Space Telescope Science Institute, which is operated by the Association of Universities for Research in Astronomy, Inc., under NASA contract NAS 5-03127 for JWST. These observations are associated with program \#1349.
Support for program \#1349 was provided by NASA through a grant from the Space Telescope Science Institute, which is operated by the Association of Universities for Research in Astronomy, Inc., under NASA contract NAS 5-03127.
\rev{The material is based upon work supported by NASA under award number 80GSFC21M0002.}

\rev{All of the data presented in this paper were obtained from the Mikulski Archive for Space Telescopes (MAST) at the Space Telescope Science Institute. The specific observations analyzed can be accessed via \dataset[10.17909/ytb0-px48]{https://doi.org/10.17909/ytb0-px48}}

%

\vspace{5mm}
\facilities{JWST/NIRISS}

\end{document}